# State capital involvement, managerial sentiment and firm innovation performance Evidence from China

Xiangtai Zuo (Shutter Zor)
Accounting College, Wuhan Textile University

**Abstract:** In recent years, more and more state-owned enterprises (SOEs) have been embedded in the restructuring and governance of private enterprises through equity participation, providing a more advantageous environment for private enterprises in financing and innovation. However, there is a lack of knowledge about the underlying mechanisms of SOE intervention on corporate innovation performance. Hence, in this study, we investigated the association of state capital intervention with innovation performance, meanwhile further investigated the potential mediating and moderating role of managerial sentiment and financing constraints, respectively, using all listed non-ST firms from 2010 to 2020 as the sample. The results revealed two main findings: 1) state capital intervention would increase innovation performance through managerial sentiment; 2) financing constraints would moderate the effect of state capital intervention on firms' innovation performance.

**1. Introduction**

State capital involvement refers to the involvement of state-owned enterprises in the production and operating activities of private enterprises through equity investment [1]. It is well known that China's economic system is not a fully market economy, with the government acting as a visible hand in charging of balance and control of the economy. According to the "Guidance Opinions of the CPC Central Committee and State Council on Deepening the Reform of State-owned Enterprises" proposed by the Central Committee of the Communist Party of China (CPC), we should actively encourage the mutual integration and mixed reform between private enterprises and state-owned enterprises and combine the advantage of both sides, with these efforts can we promote sustainable development, enhance enterprises' vitality and increase enterprise innovation capacity. In details, the mixed reform includes both allowing for private shareholders in state-owned enterprises, and the absorption of state-owned investments in private enterprises. Taking the mixed reform of Unicom in 2017 as an example. After introducing capital injection from strategic investors in the Internet, finance and other fields in form of issuance of additional directional shares, there were a 452.31% year-on-year increase in net profit, and a 45% increase in industrial Internet revenue according to 2018, as well as various innovative businesses were launched. All these suggested great benefits of the mixed reform. Participating in the mixed ownership reform helped private enterprises to improve innovation performance and innovation dynamic, as well as absorb more innovation capital [2], which further promoted its innovation output [3, 4]. Considerable studies conducted in Chinese context have also shown that the involvement of state capital contributed to the decrease in the financing constraints of companies [5, 6], due to the high credibility of the government. State capital involvement brings both more stable cash flow and more reliable financial guarantors. Thus, compared to private companies without state capital involvement, which attract investment depend on their own capacity, companies with a state capital background would have advantages of more stable cash flow and fewer financing constraints [6, 7]. Besides, because they are supported by administrative units, such as the government, they also have more access to production materials and resources [8]. All these benefits ensure managers of companies with a state capital support have more sufficient and stable resources to avoid under-investment in some innovative programs. Simultaneously, in trust to the state, banks and other institutions would be also more willing to invest in companies with state capital involvement, making it less difficult for them to raise capital, further preventing these companies from abandoning some of valuable innovation strategies.

Managers are main drivers of innovation projects, whose emotional instability could be more harmful to companies than that of ordinary employees. Theoretical and empirical researches have shown the relations between managers' overconfidence and innovative performance of companies [9-11]. For example, managers who are overconfident usually overestimate the probability of success and tend to favour highly innovative but risky projects, further increasing firm's innovation performance [12, 13], or causing irreparable losses. On the contrary, It can of course be inferred that this equally implies that a lack of self-confidence can cause similar losses to a company, as no philosophical theory would be too absolute. Numerous studies have also shown that moderately confident managers are better for business because they are more able to remain calm relative to confident managers and have sufficient decision-making power relative to less confident managers. For those less-confident and more-conservative managers, they are more likely to be cautious in investments in innovative projects. However, the results of the relations between managerial overconfidence and innovation performance is complex, as it is difficult to figure out whether it is managerial overconfidence that contributes to innovation performance, or it is overconfident managers who are preferred by more aggressive and innovative firms.

Due to the relatively low maturity of China's capital markets, some standard rational investment assumptions designed for developed countries may be not applicable in Chinese context, and there is a certain cognitive bias in the impact of managers on events [8], which is still expressed on top of the impact on firm effectiveness [14]. These factors may bring consequences of: 1) managers may abandon radical innovation plans, instead prefer some profitable projects they thought, as a result of misjudgements or just to cater for market demand for financing; 2) the success or otherwise of a strategic shift would arouse certain emotions of the manager, whether it is the joy of increasing dividends as a result of abandoning innovative projects, leading to the increase of financial performance, or the remorse at the loss of the business as a result of choosing to innovate [15], both of these emotions can never be fully measured or constructed by overconfidence alone. Hence, due to the nature of complexity in human emotions, more attention needed to pay to the association of general managers' emotions with innovative performance of companies.

In addition, lots of studies investigated this relation among high-tech firms, little has known in industry-wide samples. Therefore, the current study would focus on the following three questions in an industry-wide perspective: 1) whether state capital involvement affects general managers' sentiment; 2) whether this effect increases firms' innovation performance; and 3) whether state capital involvement affects firms' innovation performance by moderating financing constraints in addition to affecting managers' sentiment.

## 2. Theoretical analysis and research hypothesis

Firstly, the relationship between state capital intervention and firms' innovation performance is investigated; secondly, we further argue for the possible transmission role of managerial sentiment in the said relationship. and third, the transmission mechanism between managerial sentiment, which is the main innovation point of this paper, is investigated as being influenced by financing constraints. Since state capital intervention, in addition to influencing managerial sentiment, can to some extent increase the amount of investment in firms by reducing their financing constraints, thus changing the status quo of underinvestment and improving their innovation performance, the moderating role of financing constraints will also be discussed in this paper.

### 2.1. State capital involvement and innovation performance

From an input-output perspective, the hybrid reform of private enterprises with state-owned involvement can improve the innovation performance of enterprises [3, 4]. That is, firms with a background of state capital involvement would outperform those without such involvement in innovation [16]. Zeng and Guo [17], using Shanghai micro enterprise data, found that the technological innovation capacity of enterprises with state capital involvement was much higher than that of other ownership enterprises. By comparing firms with different backgrounds, Wu [18] revealed that firms with state-owned involvement but not fully state-owned had significantly stronger technological innovation capabilities than other types of firms. In a similar vein, Liu et al [19] also found that state-owned industrial enterprises had higher R&D investment and innovation performance than those private enterprises. In addition, it isn't hard to observe that enterprises with a state-owned capital background have more cash flow due to their reliance on administrative units such as the government, which makes it easier for them to solve the problem of under-investment than private enterprises. Therefore, we proposed our first hypothesis.

**Hypothesis I:** State capital involvement would improve firms' innovation performance.

### 2.2. State capital involvement and managerial sentiment

Compared to private enterprises, which are not state-owned and have to cater for the market situation, those with a state-owned background are less sensitive to changes in the market and have more access to production materials and resources [7, 8]. These advantages ensure that the managers of SOEs have sufficient and stable cash flow, so that they do not have to give their innovative projects to meet market conditions. But for private enterprises, they have to pay attention to market trends to meet market conditions and maintain a good image in the capital markets, depending on these can they make economic decisions that are more conducive to the development of their own financing. The stability of resources and the uncertainty of cash flows can lead to a lower emotional index for private managers, as they tend to be more worried about the inadequate resources. In such a situation, catering to the market's choice and abandoning innovative schemes with unstable and uncertain returns becomes a choice that can be well hedged against risk. In other words, managers of companies with a state-owned background may have more positive emotions, such as feeling more comfortable and relaxed.

**Hypothesis II:** State capital involvement would be associated with managerial sentiment.

*2.3. Managerial sentiment and corporate innovation performance*

Managerial sentiment is an abstract concept in behavioural management. Professor Mayo, following the Hawthorne experiment, suggested that human, as the most basic unit of human society, are not just "economic human" but "social human". Naturally, managers are "social beings" with complex emotions, and they are usually more competent and knowledgeable than ordinary employees. The literature has tended to focus on the impact of managerial overconfidence on innovation performance using the size of the median firm's investment in innovation in relation to the industry, suggesting that overconfident managers are more likely to engage in innovative activities [20]. However, these studies ignored the other side of overconfidence: the lack of confidence. Literally, the lack of confidence is not a negation of managers' emotions, instead it's a rigorous and self-interested fear to investment failure that could lead to a reduction in their own incentive pay. According to "short-sightedness" or principal-agent analysis, the risks of R&D investment in the existing market situation can be characterised by uncertainty and high risk, and to avoid a disastrous failure after investing many human and material resources, managers will value the investment in R&D projects in advance, during this process, the valuation model used is correlated with the available technology, and errors in valuation can direct affect their investment in innovation [21]. Hence, managers may be more conservatively and rigorous in assessments and designs if they tend to be worried about the reduction of their gains. Although it's a below-industry median investment (because managers are not confident), this rigorous innovation design may also improve the firm's innovation performance to some extent. This study therefore infers that regardless of whether managers are confident or not, their sentiment can contribute to the increase in firms' innovation performance, and that the combined sentiment is more representative of managers' true intentions when faced with an innovation strategy, leading to our hypothesis III.

**Hypothesis III:** Managerial sentiment displays a positive correlation with firm innovation performance.

*2.4. The moderating effect of financing constraints*

For private firms without or with less state involvement, due to monopoly suppression and credit discrimination, they are more likely to have financing constraints, therefore are more cautious in allocation of capital. Managers of private companies may have more frequent and greater fluctuations in management sentiment than those of state-owned companies, as they always worry about cash flow and financing. On the contrary, the benefits of stable capital, relatively easy access to finance and less changes in plans in response to market winds, all these make managers of state-owned companies have higher tolerance for risk-taking. The discussion above shows that managers of non-state enterprises tend to show more negative emotions when faced with financing constraints than managers of state-owned enterprises, i.e., they score lower on the emotion indicator. This leads to different views on the need for innovation decisions by managers of the two types of firms, i.e. that financing constraints moderate the interaction between state capital involvement and managerial sentiment and firm innovation performance. As a result of the above evidence, the fourth hypotheses of this paper can be inferred. And also the research framework of this paper can be found according to Figure 1.

**Hypothesis IV:** Financing constraints would moderate the relationship between state capital involvement and firm innovation performance.

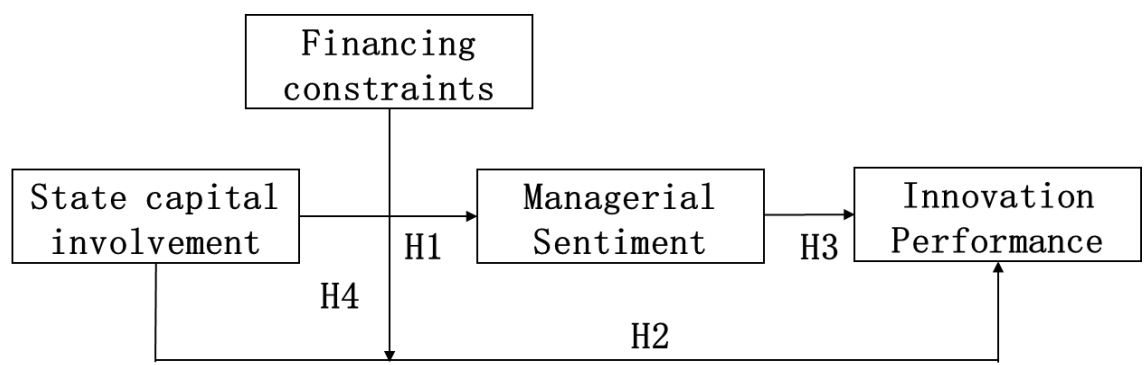

Figure 1. Conceptual model

## 3. Research design

*3.1. Data and samples*

In this study, all listed non-ST companies from 2010-2020 were selected from the Csmar[1] database as the study sample, and the samples were cross-sectionally merged by stock code and year to remove duplicate redundancy, resulting in a total of 7,294 samples. All analyses in our study were conducted using Stata 16.0.

*3.2. Definition of variables*
3.2.1. Innovation Performance (*INNO*)

Innovation performance generally includes innovation inputs and innovation outputs, but effective innovation outputs are easier to obtain and define, so referring to Zhu et al [5], the natural logarithm of the number of lagged one-period patent

---
[1] Csmar, which has one of the most comprehensive databases of listed companies in China, can be found at: http://cndata1.csmar.com/.

applications of listed companies was used to measure companies' innovation performance.

3.2.2. State capital involvement (*NATION*)

State-owned involvement here refers to the shareholding of Chinese state-owned enterprises in private enterprises, and it was measured by the shareholding ratio of Chinese state-owned shareholders among the top 10 shareholders of private enterprises.

3.2.3. Managerial Sentiment (*SENT*)

The content of the emotion lexicon was constructed according to Yao et al [22], then positive and negative sentiments were calculated separately by intersecting with managerial disclosures, so that the proxy variable for managerial sentiment (SENT) = managerial positive sentiments - managerial negative sentiments.

For the construction of the sentiment lexicon, the following word frequency method with a penalty mechanism was used to calculate the words in the initial lexicon that conformed to the annual reporting conventions and were adjusted for the emotional lexicon.

$$positive = \frac{w_{n,N}}{\sum_n w_{n,N}} \times \frac{1}{1 + \frac{w_{n,P}}{\sum_n w_{n,P}}} \qquad (1)$$

Where, $W_{n,N}$ is the number of occurrences of optional positive words $n$ in the set $N$ of annual reports with positive returns in the vocabulary of the annual report corpus from the Integrated Chinese Lexical Analysis System of the Chinese Academy of Sciences and others. $W_{n,P}$ is the number of occurrences of the optional positive words $n$ in the set $N$ of annual reports with negative meaning. It should be noted that the score of positive words in the annual report increases with the left half of the right-hand side of the equation. With the introduction of a penalty mechanism for the right-hand side of the equation, the misclassification of positive word order can be reduced, and the words can be filtered and sorted appropriately. The same principle was applied to the screening of negative words. In this way, a more reliable and reproducible sentiment lexicon can be obtained with a consistent corpus and method, and the sentiment lexicon section (words in no particular order) was given in Table 1.

Table 1. Sentiment Dictionary (partial)

| Dictionary type | Part of the content |
| --- | --- |
| Positive | Top 100, Patriotic, Hobby, Settlement, Safe and Secure, Hundred Flowers, Top of the List, Inclusive ...... |
| Negative | Settling for the status quo, concealed, marginalised, backtracking, deviating, barriers, on the verge, not up to scratch ...... |

After obtaining an objective and accurate sentiment lexicon, the text (*basicmess*) of the basic information in the annual reports of listed companies was intersected and filtered through Stata's text analysis function, in which positive words (*positive*) added to the positive score and negative words (*negative*) added to the negative score, and finally aggregated into a comprehensive sentiment indicator for managers, calculated as follows.

$$positive_{it} = \begin{cases} 1, & \text{if } positive \cap basicmess = true; \\ 0, & \text{if } positive \cap basicmess = false. \end{cases}$$
$$negative_{it} = \begin{cases} 1, & \text{if } negative \cap basicmess = true; \\ 0, & \text{if } negative \cap basicmess = false. \end{cases} \quad (2)$$

$$Score_{positive_{it}} = \sum positive_{it}$$
$$Score_{negative_{it}} = \sum negative_{it} \quad (3)$$

$$SENT_{it} = Score_{positive_{it}} - Score_{negative_{it}} \quad (4)$$

where $positive_{it}$ represented the total positive sentiment score of a company for a given year calculated by equation (2) and $negative_{it}$ represented the total negative sentiment score of a company for a given year calculated by equation (2).

3.2.4. Financing constraints (*SA*)

Referring to Ye's [6] definition, the more conventional SA index was used to define the size of the financing constraint faced by the firm, with a larger SA indicating a larger financing constraint. *SA = 0.434\*SIZE2 - 0.737\*SIZE - 0.04\*AGE.*

3.2.5. Other control variables

In order to make the relationships between the three types of variables mentioned above more salient, some control variables were introduced with reference to previous studies (see Table 2).

Table 2. Other control variables

| Variable name | symbol | Calculation method |
| --- | --- | --- |
| Leverage | LEV | Total liabilities/total assets |
| Return on total assets | ROA | Net Profit / Total Assets |
| Age | AGE | Year of sample - year of company establishment |
| Size | SIZE | Natural logarithm of total assets |
| Shareholding concentration | CR5 | Sum of the top five shareholders' shareholdings |
| Degree of shareholding balance | Z | Shareholding of the second to fifth largest shareholders / Shareholding of the largest shareholder |

*3.3. Estimation models*

To begin with, in order to verify the mediation effect between state capital involvement, managerial sentiment and firm innovation performance, we adopted the conduction mechanism test suggested by Lin et al [23] and constructed the following model.

$$SENT = \alpha_0 + \alpha_1 NATION + \alpha_2 LEV + \alpha_3 ROA + \alpha_4 AGE + \alpha_5 SIZE + \alpha_6 CR5 + \alpha_7 Z + \varepsilon \quad (5)$$

$$INNO = \alpha_0 + \alpha_1 NATION + \alpha_2 LEV + \alpha_3 ROA + \alpha_4 AGE + \alpha_5 SIZE + \alpha_6 CR5 + \alpha_7 Z + \varepsilon \quad (6)$$

$$INNO = \alpha_0 + \alpha_1 SENT + \alpha_2 LEV + \alpha_3 ROA + \alpha_4 AGE + \alpha_5 SIZE + \alpha_6 CR5 + \alpha_7 Z + \varepsilon \quad (7)$$

$$INNO = \alpha_0 + \alpha_1 NATION + \alpha_2 SENT + \alpha_3 LEV + \alpha_4 ROA + \alpha_5 AGE + \alpha_6 SIZE + \alpha_7 CR5 + \alpha_8 Z + \varepsilon \quad (8)$$

The three equations mentioned above examined H1, H2 and H3 respectively and tested the conducting effect of managerial sentiment. Later, to test H4, the moderating effect of financing constraints (*SA*) was introduced and model (8) was developed.

$$INNO = \beta_0 + \beta_1 NATION \times SA + \beta_2 NATION + \beta_3 SA + \beta_4 SENT + \sum Control + \varepsilon \quad (9)$$

where, $\alpha_0$ was the intercept (constant), $\alpha_n$ (n>0) was the coefficient of the paths, *Control* was the control variable and $\varepsilon$ was the residual.

## 4. Result

*4.1. Descriptive statistics and correlation analysis*

Table3. Data characteristics and correlation of study variables

| Variables | N | Min | Max | Mean | Std | INNO | SENT | NATION | SA |
|---|---|---|---|---|---|---|---|---|---|
| INNO | 7294 | 0 | 10.883 | 2.626 | 2.383 | 1 | | | |
| SENT | 7294 | -1 | 100 | 10.050 | 5.161 | 0.066*** | 1 | | |
| NATION | 7294 | 0 | 1 | 0.056 | 0.106 | 0.115*** | 0.085*** | 1 | |
| SA | 7294 | 0.832 | 12.132 | 3.497 | 1.175 | 0.244*** | 0.081*** | 0.351*** | 1 |

Note: *, ** and *** indicate significant at the 10%, 5% and 1% levels of significance respectively.

As shown in the above table, the number of patent applications in the industry-wide sample between 2010 and 2020 was approximately 13.818. Overall, state capital involvement and managerial sentiment in all the samples differed in terms of mean, minimum and maximum values, indicating that there may be large differences between different groups of companies, i.e., the companies in study samples were representative and conclusions we drawn were generalizable. Other variables fluctuated within an acceptable range.

*4.2. Regression results and analyses*

To investigate the conducting effect of managerial sentiment and the moderating effect of financing constraints, regressions analyses were run on the sample data, and models (1)-(5) in Table 4 corresponded to the regression results of equations (5)-(9), respectively.

Table 4. The conducting effect of managerial sentiment and the moderating effect of financing constraints

| VARIABLES | (1) SENT | (2) INNO | (3) INNO | (4) INNO | (5) INNO |
|---|---|---|---|---|---|
| SENT | | | 0.024*** | 0.024*** | 0.023*** |
| | | | (4.732) | (4.626) | (4.612) |
| NATION | 3.265*** | 0.479* | | 0.403 | 3.848*** |
| | (5.413) | (1.825) | | (1.532) | (5.088) |
| SA | | | | | -0.389 |
| | | | | | (-0.959) |
| NATION×SA | | | | | -0.750*** |
| | | | | | (-4.568) |
| LEV | -0.333 | -0.177 | -0.169 | -0.169 | -0.170 |
| | (-0.808) | (-0.984) | (-0.946) | (-0.942) | (-0.949) |
| ROA | -5.126*** | -3.322*** | -3.231*** | -3.202*** | -3.089*** |
| | (-5.970) | (-8.886) | (-8.642) | (-8.554) | (-8.253) |
| AGE | -0.016 | 0.024*** | 0.024*** | 0.024*** | 0.003 |
| | (-1.426) | (4.975) | (5.051) | (5.059) | (0.191) |
| SIZE | 0.342*** | 0.542*** | 0.549*** | 0.534*** | 1.084** |
| | (4.610) | (16.788) | (17.727) | (16.538) | (2.363) |
| CR5 | -0.006*** | 0.007*** | 0.007*** | 0.007*** | 0.007*** |
| | (-6.967) | (17.850) | (18.254) | (18.192) | (18.658) |
| Z | 0.088*** | 0.065*** | 0.062*** | 0.063*** | 0.059*** |
| | (3.757) | (6.382) | (6.078) | (6.181) | (5.830) |
| Constant | 4.091*** | -10.777*** | -11.160*** | -10.874*** | -21.073** |
| | (2.697) | (-16.318) | (-17.634) | (-16.479) | (-2.566) |
| Observations | 7,294 | 7,294 | 7,294 | 7,294 | 7,294 |
| R-squared | 0.025 | 0.132 | 0.134 | 0.135 | 0.140 |
| Adj_r2 | 0.139 | 0.131 | 0.139 | 0.139 | 0.139 |
| F | 118.3 | 158.4 | 118.3 | 118.3 | 118.3 |

Note: *, ** and *** indicate significant at the 10%, 5% and 1% levels of significance respectively, with the t-statistic in parentheses.

The results revealed that in model (1), Chinese capital intervention was positively significant associated with managerial sentiment, which supported our H2. In model (2), managerial sentiment was positively significant associated with firms' innovation, which supported H1. Furthermore, the mediating role of managerial sentiment between state capital involvement and the innovation performance was tested using model (1), (2), (4) and the stepwise test of the mediation effect proposed by Baron and Kenny [24], the results was shown in Figure 2.

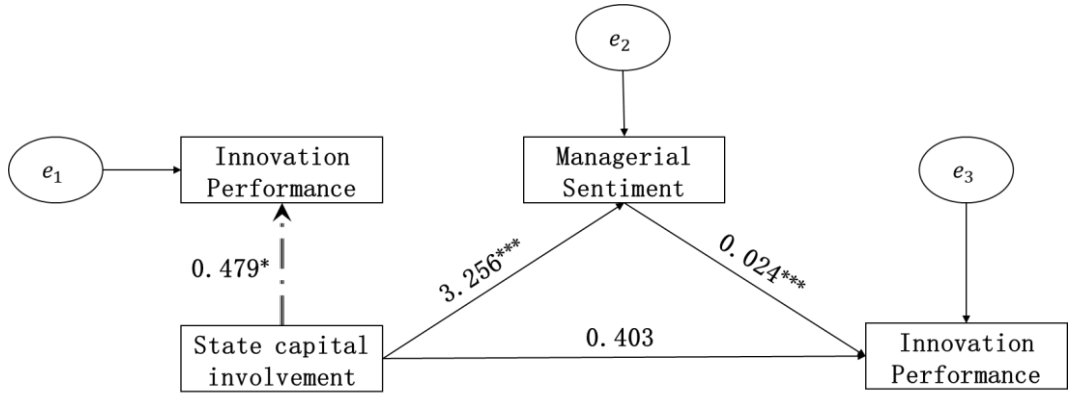

Figure 2. Illustration of the intermediary effect

Referring to the findings of Wen [25] and others [24], it is easy to find that the coefficients of the independent variables NATION and SENT are significant in model (1) and model (2), but in model (4) when Chinese capital intervenes the coefficient before the variable NATION is not significant, which means that managerial sentiment SENT shows a more significant influence in such a transmission path.

Model (5) examined the moderating effect of the financing constraint (*SA*) with a coefficient of -0.750 at the 1% level of significance. Model (4) and model (5) correspond to regression models (8) and (9) respectively, in which the magnitude and direction of the effect of the inclusion of *SA* on firms' innovation performance can be obtained by taking the partial derivative of *NATION* in these two regression models.

$$\frac{\partial INNO}{\partial NATION} = \alpha_1 \qquad (10)$$

$$\frac{\partial INNO}{\partial NATION} = \beta_1 SA + \beta_2 \qquad (11)$$

From Table 4, $\alpha_1$ and $\beta_2$ are greater than zero, $\beta_1$ is less than zero and *SA* is greater than zero according to Table 3, so it is not difficult to conclude that the presence of financing constraints (*SA*) will significantly and negatively moderate the impact of state capital intervention on firms' innovation performance, i.e., hypothesis H4 is proved.

*4.2. Stability tests*

In addition, to test the robustness of H1, we used a between-group t-test to test whether there was a difference in innovation performance among firms (with vs without state capital involvement), and the results were shown in Table 5. Moreover, the dummy variable of R&D investment indicted the percentage of operating revenue while the dummy variable of the number of R&D staff referred to managerial overconfidence (the ratio which was greater than the median of the sample in the year was coded as 1, otherwise was coded as 0). The results were shown in Tables 6 and 7, respectively.

Table 5. T-test between groups

| Group | Observations | Mean | T-test |
|---|---|---|---|
| *NATION*(0) | 4850 | 2.410 | -10.9953*** |
| *NATION*(1) | 2444 | 3.055 | |

The results of the t-test showed the innovation performance of firms with state involvement was significantly higher than those firms without state involvement i.e. the robustness of the H1 hypothesis was verified.

Table 6. Robustness test for replacement variables I (R&D input ratio)

| VARIABLES | (6) SENT | (7) INNO | (8) INNO | (9) INNO |
|---|---|---|---|---|
| SENT |  | 0.674*** | 0.671*** | 0.663*** |
|  |  | (12.718) | (12.630) | (12.502) |
| NATION | 0.294*** |  | 0.282 | 3.323*** |
|  | (5.137) |  | (1.083) | (4.425) |
| SA |  |  |  | -0.575 |
|  |  |  |  | (-1.430) |
| NATION×SA |  |  |  | -0.655*** |
|  |  |  |  | (-4.017) |
| LEV | -0.513*** | 0.169 | 0.167 | 0.164 |
|  | (-13.097) | (0.939) | (0.932) | (0.913) |
| ROA | -0.504*** | -3.005*** | -2.984*** | -2.869*** |
|  | (-6.177) | (-8.113) | (-8.047) | (-7.732) |
| AGE | 0.001 | 0.023*** | 0.023*** | -0.005 |
|  | (0.956) | (4.880) | (4.888) | (-0.295) |
| SIZE | -0.050*** | 0.586*** | 0.576*** | 1.328*** |
|  | (-7.072) | (19.130) | (17.955) | (2.920) |
| CR5 | -0.000** | 0.007*** | 0.007*** | 0.007*** |
|  | (-2.240) | (18.411) | (18.368) | (18.794) |
| Z | 0.016*** | 0.054*** | 0.054*** | 0.051*** |
|  | (7.108) | (5.310) | (5.380) | (5.049) |
| Constant | 1.725*** | -12.140*** | -11.935*** | -25.739*** |
|  | (11.971) | (-19.208) | (-18.089) | (-3.159) |
| Observations | 7,294 | 7,294 | 7,294 | 7,294 |
| R-squared | 0.062 | 0.151 | 0.151 | 0.155 |
| Adj_r2 | 0.154 | 0.154 | 0.154 | 0.154 |
| F | 133.9 | 133.9 | 133.9 | 133.9 |

Note: *, ** and *** indicate significant at the 10%, 5% and 1% levels of significance respectively, with the t-statistic in parentheses.

Table 7. Robustness tests for replacement variables II (R&D staff ratio)

| VARIABLES | (10) SENT | (11) INNO | (12) INNO | (13) INNO |
|---|---|---|---|---|
| SENT |  | 0.199*** | 0.196*** | 0.208*** |
|  |  | (3.524) | (3.478) | (3.690) |
| NATION | 0.122** |  | 0.456* | 3.931*** |

|  | (2.240) |  | (1.735) | (5.196) |
|---|---|---|---|---|
| SA |  |  |  | -0.427 |
|  |  |  |  | (-1.053) |
| NATION×SA |  |  |  | -0.756*** |
|  |  |  |  | (-4.601) |
| LEV | -0.218*** | -0.134 | -0.134 | -0.132 |
|  | (-5.863) | (-0.747) | (-0.744) | (-0.736) |
| ROA | -0.232*** | -3.312*** | -3.277*** | -3.157*** |
|  | (-2.992) | (-8.872) | (-8.765) | (-8.444) |
| AGE | -0.015*** | 0.027*** | 0.027*** | 0.004 |
|  | (-15.364) | (5.514) | (5.517) | (0.258) |
| SIZE | -0.061*** | 0.571*** | 0.554*** | 1.150** |
|  | (-9.064) | (18.389) | (17.074) | (2.504) |
| CR5 | -0.001*** | 0.007*** | 0.007*** | 0.007*** |
|  | (-6.642) | (18.141) | (18.079) | (18.570) |
| Z | 0.008*** | 0.062*** | 0.064*** | 0.060*** |
|  | (3.682) | (6.113) | (6.231) | (5.862) |
| Constant | 2.316*** | -11.561*** | -11.232*** | -22.276*** |
|  | (16.917) | (-17.906) | (-16.695) | (-2.708) |
| Observations | 7,294 | 7,294 | 7,294 | 7,294 |
| R-squared | 0.076 | 0.133 | 0.133 | 0.139 |
| Adj_r2 | 0.138 | 0.138 | 0.138 | 0.138 |
| F | 117.4 | 117.4 | 117.4 | 117.4 |

Note: *, ** and *** indicate significant at the 10%, 5% and 1% levels of significance respectively, with the t-statistic in parentheses.

As shown in Tables 6 and 7, after we replaced managerial sentiment with different indicators of managerial overconfidence, relationships between variables were still significant and in line with our study results. Overall, the robustness tests revealed following results: 1. State capital involvement still contributed to the rise in managerial sentiment (models (6) and (10)), suggesting the robustness of H2; 2. Managerial sentiment significantly positive affected the level of innovation performance of the firm (models (7) and (11)), suggesting the robustness of H3; 3. The moderating effect of financing constraints (SA) after replacing the indictors of managerial sentiment was still negative and significant (models (8), (9) and (12), (13)), demonstrating the robustness of H4.

Finally, we tested the effect of SA on managerial sentiment and innovation performance under different levels of financing constraints. The median of the financing constraints (SA) was used as a cut-off score, with firms below the median being those with low financing constraints (coded as 0), vice versa (coded as 1).

Table 8. Impact of state capital involvement under different financing constraints

|  | (14) | (15) | (16) | (17) |
|---|---|---|---|---|
|  | SA | | | |

|  | Low | High | Low | High |
|---|---|---|---|---|
| VARIABLES | SENT | SENT | INNO | INNO |
| NATION | 5.631*** | 2.765*** | 1.836*** | 0.310 |
|  | (4.918) | (3.757) | (4.034) | (0.906) |
| LEV | -0.396 | -0.029 | -0.304 | 0.072 |
|  | (-0.659) | (-0.051) | (-1.272) | (0.273) |
| ROA | -6.899*** | -2.891** | -4.025*** | -2.143*** |
|  | (-5.694) | (-2.349) | (-8.355) | (-3.745) |
| AGE | -0.044*** | 0.003 | 0.007 | 0.040*** |
|  | (-2.814) | (0.190) | (1.094) | (5.306) |
| SIZE | 0.793*** | 0.252** | 0.704*** | 0.417*** |
|  | (3.786) | (2.081) | (8.457) | (7.412) |
| CR5 | -0.006*** | -0.006*** | 0.009*** | 0.005*** |
|  | (-4.722) | (-4.890) | (17.920) | (8.050) |
| Z | 0.060* | 0.109*** | 0.060*** | 0.051*** |
|  | (1.831) | (3.159) | (4.672) | (3.198) |
| Constant | -4.818 | 5.657** | -14.268*** | -7.850*** |
|  | (-1.133) | (2.221) | (-8.441) | (-6.626) |
| Observations | 3,635 | 3,659 | 3,635 | 3,659 |
| R-squared | 0.030 | 0.019 | 0.156 | 0.066 |
| Adj_r2 | 0.0278 | 0.0278 | 0.0278 | 0.0278 |
| F | 15.83 | 15.83 | 15.83 | 15.83 |

Note: *, ** and *** indicate significant at the 10%, 5% and 1% levels of significance respectively, with the t-statistic in parentheses.

The results revealed that whether the financing constraint is high or low, the involvement of state capital is conducive to an increase in the sentiment of corporate managers, as state capital involvement is a positive sign for corporate managers, and with the support of state capital, corporate managers will show higher sentiment towards their own behaviors, as in this case managers no longer need to worry about the adequacy of funding and have a higher tolerance for error. And the impact of state capital involvement on firms' innovation performance was negatively affected by the financing constraint. From the result of model (16), we could find that state capital involvement was significantly positive related to firms' innovation performance under low financing constraint. However, model (17) showed no non-significant relationship between state capital involvement and firms' innovation performance when the financing constraint was high. Overall, these results reflected the negative moderating effect of financing constraint (SA), again validating the robustness of H4.

## 4. Discussion and recommendations

Previous research has only explored the impact of equity structure on firm innovation performance in terms of managerial overconfidence, and the sample was limited to high-technology firms, resulting in their findings lacking generalizability. The current study expanded the measurement of managerial sentiment and investigated relationships between state capital involvement, managerial sentiment and corporate innovation performance, using a sample of non-ST companies listed in China from 2010 to 2020. Our results revealed following findings at least among all studied listed companies: 1. State capital involvement has a significantly positive effect on corporate managerial sentiment; 2. State capital involvement has a significantly positive effect on corporate innovation performance; 3. Managerial sentiment mediates the relation between state involvement and firms' innovation performance; 4. Financing constraints negatively moderate the positive relationship between state capital involvement and firms' innovation performance.

The current study defined a new measurement of managerial sentiment and provided a reference for the mechanism of how state capital improved the innovative performance of firms. Despite these theoretical contributions, our findings may also have more implications for practice. First, managers should be appropriately 'irrational', which requires companies to provide incentives and penalties that are not entirely dependent on the financial performance. These measures may help to reduce the agency problems and 'short-sightedness' caused by 'fully rational' behaviors of managers, so that they can make innovative decisions that are conducive to the long-term development of the company, and ensure their companies at a reasonable level of innovative performance, even when they have sufficient funds to meet market conditions. Second, the involvement of state capital can help to mitigate the negative impact of managers' negative emotions on the innovation performance of companies, because of the reliance on state capital and financing support. Hence, the state government may support some prospective new technology industries, especially through state ownership, giving them enough support to reduce the risk of innovation caused by over- reliant on the market.